\documentclass[twocolumn,aps,prb,showpacs,preprintnumbers,amsmath,amssymb,psfrac]{revtex4}

\usepackage{epsfig}
\usepackage{graphicx}
\usepackage{dcolumn}
\usepackage{bm}

\renewcommand{\dag}{^{\dagger}}

\def\gapp{\lower.35em\hbox{$\stackrel{\textstyle>}{\sim}$}} 
\def\lapp{\lower.35em\hbox{$\stackrel{\textstyle<}{\sim}$}} 

\begin{document}
%

\title{
Assisted hopping in the Anderson impurity model: A flow equation study
}
\author{T. Stauber and F. Guinea}
\affiliation{Instituto de Ciencia de Materiales de Madrid, CSIC, Cantoblanco, E-28049 Madrid, Spain.
}
\date{\today}
\begin{abstract}
We investigate the effect of assisted hopping in the 
Anderson impurity model. We use the flow equation method,
which, by means of unitary transformations,
generates a sequence of Hamiltonians in order to
eliminate the assisted hopping terms.
This approach yields a renormalized on-site energy $\epsilon_d^*$, 
a renormalized correlation energy $U^*$, 
and other terms, which include pair hopping.
For some parameter values, the initial Hamiltonian flows
towards an attractive Anderson model.
We argue that this result implies a tendency towards local pairing
fluctuations.
\end{abstract}
%
%
%
%
\maketitle
\section{Introduction.}
The properties of magnetic impurities in metals, and of small dots
attached to metallic leads share many characteristics, due to the
similar role played by electron electron interactions.
The electrostatic repulsion leads to the formation of
local moments\cite{K64,A61}, which are quenched 
at sufficiently low temperatures.
Quantum dots are more complicated structures\cite{A00,ABG02} 
and an effective Kondo Hamiltonian can only be defined in the 
limit of large spacing between the electronic levels within the dot.
Then, the main physical processes at low temperatures are due
to the changes of the occupancy of the level closest to the
Fermi energy of the leads. Even if we assume this restriction, 
terms beyond the Kondo Hamiltonian can arise due to the 
finite extension and inhomogeneities of the electron states
inside a mesoscopic quantum dot. The simplest of these additional
interactions is an assisted hopping term, extensively studied
in relation to bulk correlated systems\cite{H93,HM91}. It can be shown
that, within a systematic expansion in the intradot conductance, 
the leading correction to the intradot capacitance, or averaged
Coulomb interaction, has such a form\cite{G02}. The inhomogeneous
screening and effective ^^ ^^ orthogonality catastrophe " which give
rise to this term also lead to interesting non equilibrium effects
in the metallic limit, when the level spacing is much less
than the temperature\cite{UG91,Betal00}.

The present work analyzes the model of a correlated impurity, or
quantum dot, in the limit where the low temperature
physical properties are determined by a single electronic state within
the dot, using the flow equations first proposed in Refs. \onlinecite{W94} and \onlinecite{GW94}. 
The method transforms the initial Hamiltonian into a family
of related Hamiltonians by means of a sequence of unitary
transformations. If adequately chosen, these transformations
lead to simplified models, in a similar way to a Renormalization
Group transformation. The method has proven useful for a variety
of problems\cite{KM94,LM96}, and  it has also been used to obtain the
Kondo Hamiltonian from the Anderson model\cite{KM94}.
 
In the following, we will apply the flow equation technique to
the Anderson Hamiltonian with assisted hopping. This problem
has been analyzed in\cite{G02} using mean field techniques, but
there are no more accurate studies in the literature. Thus,
besides the intrinsic interest of the model, this work
provides a test case where the flow equation method is used
to obtain information not available by other techniques.

The next section presents the model, and describes how
the flow equation method is applied. 
The results are discussed in Section III. Finally, Section IV 
analyzes the main properties of the model, as derived by this method.
\section{The model and flow equations}
\subsection{The model}
We will study the Hamiltonian:
\begin{widetext}
\begin{eqnarray}
{\cal H} &= &{\cal H}_K + {\cal H}_{imp}
+ {\cal H}_{hyb} + {\cal H}_{assisted} \nonumber \\
{\cal H}_K &= &\sum_{k,s}\epsilon_k^0 c_{k,s}\dag c_{k,s} \nonumber \\ 
{\cal H}_{imp} &= &
\epsilon_d^0 n_d+U^0n_{d,+}n_{d,-} \nonumber \\
{\cal H}_{hyb} &= &
\sum_{k,s}V_k^0\left( c_{k,s}\dag d_s+d_s\dag c_{k,s} \right)
\nonumber \\
{\cal H}_{assisted} &= &\sum_{k,s}W_k^0n_{d,s}\left( c_{k,-s}
\dag d_{-s}+d_{-s}\dag c_{k,-s}\right)
\label{hamil}
\end{eqnarray}
\end{widetext}
where ${\cal H}_A = {\cal H}_K + {\cal H}_{imp}
+ {\cal H}_{hyb}$ is the Anderson Hamiltonian, and 
the assisted
hopping terms are included in ${\cal H}_{assisted}$.
We have also defined
$n_{d,s}=d_s\dag d_s$ and $n_d=n_{d,+}+n_{d,-}$.
\subsection{The flow equation method}
We now perform a sequence of infinitesimal unitary 
transformations:
\begin{align}
\partial_\ell {\cal H}=[\eta,{\cal H}]
\label{transformation}
\end{align}
These transformations are characterized 
by the anti-Hermitian generator $\eta=-\eta\dag$. 
The parameters of the initial Hamiltonians become 
functions of the flow parameter $\ell$ and, if there 
are no upper indices on the parameters, it will be implied 
that they depend on $\ell$.

Our main goal is to reduce the initial Hamiltonian,
Eq. (\ref{hamil}) to another whose physical properties
are well understood. We assume that this is achieved if
the assisted hopping term is made to flow to zero. Hence,
we impose the requirement
\begin{equation} 
\lim_{\ell \rightarrow \infty} {\cal H}_{assisted} ( \ell )
\rightarrow 0\quad.
\label{limit}
\end{equation}
This condition is satisfied if $\eta = [ {\cal H}_K , 
{\cal H}_{assisted} ]$. More generally, we will employ   
the following generator:
\begin{equation}
\label{generator}
\eta = \sum_{k,s}\eta_kn_{d,s}\left(c_{k,-s}\dag 
d_{-s}-d_{-s}\dag c_{k,-s}\right)
\end{equation}
This expression differs from the generator used to cancel
the hybridization with the band, $V_k^0$ in Eq. (\ref{hamil}), in that it
contains an additional operator related to the localized
orbital, $n_{d,s}$.
It is interesting to note that for the analytic treatment of the flow equations the 
explicit expression of the parameters $\eta_k$ is not needed. 
We only require that Eq. (\ref{limit}) holds.

The physical properties of the impurity
are determined by the flow of the on-site energy 
$\epsilon_d^0$ and the correlation energy $U^0$  as 
$\ell \rightarrow \infty$. Notice that the one particle hybridization in the
Hamiltonian, Eq. (\ref{hamil}), is not changed during
the flow, i.e., 
\begin{align}
\partial_\ell {\cal H}_{hyb} = 0\quad.
\label{hyb_flow}
\end{align}
This is a consequence of our choice of the generator given in Eq. (\ref{generator}). 

The commutator $[\eta,{\cal H}]$ generates new interactions
which have to be included in the Hamiltonian
${\cal H} ( \ell )$, which, in turn, will lead to other
interactions. This infinite hierarchy needs to be decoupled
in order for the method to be useful. This point is carefully
discussed in\cite{KM94}, and we will follow the same procedure.
We do not include the new operators in the flow, although they
will appear in the fixed point Hamiltonian. The role of
these ``marginal'' operators will be discussed in  
section IIIC.

The resulting flow equations thus take on the simple form
\begin{align}
\partial_\ell \epsilon_d&=2\sum_k
\eta_k(V_k^0+W_k)n_k^0\quad,\label{FlowD}\\
\partial_\ell U&=-4\sum_k\eta_k(V_k^0+W_k)\quad,\label{FlowU}\\
\partial_\ell W_k&=-(\epsilon_k^0-
\epsilon_d-U)\eta_k\quad,
\label{FlowW}
\end{align}
where we have introduced the occupation number 
$n_k^0=(e^{\beta\epsilon_k^0}+1)^{-1}$, which arises from 
normal ordering the neglected operators\cite{FootCP}. 
The  procedure used here 
explicitly breaks particle-hole symmetry. 
Note that there is no renormalization of the energies 
of the conduction electrons, $\epsilon_k^0$, 
which is a typical feature of impurity problems\cite{KM97}.

From Eq. (\ref{FlowW}), we obtain the 
general relation between the flow of the assisted 
hopping amplitudes and the generator of the infinitesimal 
transformations,
\begin{align}
\label{etaK}
\eta_k=-\frac{\partial_\ell W_k}{\epsilon_k^0-\epsilon_d-U}\quad.
\end{align}
Inserting this expression in Eqs. 
(\ref{FlowD}) and (\ref{FlowU})  we obtain
\begin{align}
\partial_\ell \epsilon_d&=-
2\sum_k\frac{(V_k^0+W_k)\partial_\ell W_k}
{\epsilon_k^0-\epsilon_d-U}n_k^0\quad,\label{flowE}\\
\partial_\ell U&=4\sum_k\frac{(V_k^0+W_k)
\partial_\ell W_k}{\epsilon_k^0-\epsilon_d-U}\quad.
\label{flowU}
\end{align}
Note that the above equations imply that
\begin{align}
\label{UniF}
\epsilon_d^*-\epsilon_d^0&=F(\epsilon_d^0+U^0)\quad,\\
U^*-U^0&=G(\epsilon_d^0+U^0)\label{UniG}
\quad,
\end{align}
where $\epsilon_d^*\equiv\epsilon_d(\ell=\infty)$ and 
$U^*\equiv U(\ell=\infty)$, and $F$ and $G$ denote universal functions. These functions only depend on the coupling constants $V_k^0$ and $W_k^0$. 

The above flow equations can also be treated semi-analytically by
substituting the $\ell$-dependent parameters 
$\epsilon_d$ and $U$ in the denominator by their fixed point values 
$\epsilon_d^*$ and 
$U^*$. This strategy has proven to be successful in several related models.\cite{KM94,Cri97}
We can then integrate the equations (\ref{flowE}) and (\ref{flowU})
from $\ell = 0$ to $\ell = \infty$, using the constraint
that $\lim_{\ell \rightarrow \infty} W_k
\rightarrow 0$.

We define
\begin{align}
J(\epsilon)&\equiv\sum_k\left(2|V_k^0W_k^0|-W_k^0W_k^0\right)
\delta(\epsilon-\epsilon_k^0)\quad,
\label{spectral}
\end{align}
where we assume that the hybridization $V_k^0$ and the assisted
hopping amplitude $W_k^0$ differ in sign\cite{H93}.
The spectral function is positive for the physically 
relevant parameter regime, where the hybridization 
amplitude is larger or comparable to the assisted hopping amplitude. 
With the above definition we obtain the final result
\begin{align}
\epsilon_d^*-\epsilon_d^0&=-\int d\epsilon
\frac{J(\epsilon)}{\epsilon-\epsilon_d^*-U^*}
f(\epsilon)\quad,\label{Intepsilon}\\
U^*-U^0&=2\int d\epsilon\frac{J(\epsilon)}{\epsilon-\epsilon_d^*-U^*}
\quad,\label{IntU}
\end{align}
where we have defined the Fermi function 
$f(\epsilon)\equiv(e^{\beta\epsilon}+1)^{-1}$.\cite{FootCP}

\section{Results}
\subsection{Effective parameters}
In order to make the analysis more quantitative, we need
to specify the density of states in the surrounding medium, described
by the function $J ( \epsilon )$,
Eq. (\ref{spectral}). 
The width of the conduction band
in the leads is
$2D$, and the Fermi energy is $\epsilon_F = 0$. We also define
\begin{eqnarray}
V &\equiv &\sqrt{\sum_k(V_k^0)^2} \nonumber \quad,\\
W &\equiv &\sqrt{\sum_k(W_k^0)^2} \quad,\\
\Gamma &\equiv &\pi\rho(\epsilon_F)(2|V_{k_F}^0
W_{k_F}^0|-(W_{k_F}^0)^2)=\pi J(\epsilon_F)\quad.
\nonumber
\end{eqnarray}
We further assume that both coupling functions, $V_k^0$ and $W_k^0$, are semi-elliptic. This choice resembles a realistic environment and facilitates the computation.

To obtain the universal functions $F$ and $G$ from Eqs. (\ref{UniF}) and (\ref{UniG}), we numerically integrate the flow equations  (\ref{FlowD} - \ref{FlowW}) with 
\begin{align}
\eta_k=(\epsilon_k^0-\epsilon_d-U)W_k\quad.
 \end{align}
This choice guarantees that $W_k$ vanishes exponentially for $\epsilon_k^0\neq\epsilon_d^*+U^*$ and algebraically for $\epsilon_k^0=\epsilon_d^*+U^*$.\cite{KM94} By this, we introduce an energy-scale dependent decoupling scheme with respect to the renormalized energy of the double occupancy of the dot, $\epsilon_d^*+U^*$.

We now turn to the semi-analytical solution.
The spectral function of Eq. (\ref{spectral}) reads
\begin{align}
\label{SemiCircular}
J(\epsilon)&\equiv\Theta(D^2-\epsilon^2)
\frac{\Gamma}{\pi D}\sqrt{D^2-\epsilon^2}\quad,
\end{align}
with the resonance $\Gamma=2(2VW-W^2)/D$. 
Evaluating Eqs. (\ref{Intepsilon}) and (\ref{IntU}) 
we have to distinguish two cases:
\begin{widetext}
For $|\epsilon_d^*+U^*|<D$, we 
obtain the following self-consistent equations:
\begin{align}
\label{ErenormalizedLLD}
\epsilon_d^*-\epsilon_d^0
&=\frac{\Gamma}{2D}\left[(\epsilon_d^*+U^*)-\frac{2}{\pi}D
+\frac{2}{\pi}\sqrt{D^2-(\epsilon_d^*+U^*)^2}
\ln\frac{|\epsilon_d^*+U^*|}{D-\sqrt{D^2-
(\epsilon_d^*+U^*)^2}}\right]\quad,\\
\label{UrenormalizedLLD}
U^*-U^0&=-\frac{2\Gamma}{D}(\epsilon_d^*+U^*)
\end{align}

For $|\epsilon_d^*+U^*|>D$, we obtain:
\begin{align}
\label{ErenormalizedGGD}
\epsilon_d^*-\epsilon_d^0&=\frac{\Gamma}{2D}
\left[(\epsilon_d^*+U^*)-\frac{2}{\pi}D
-\text{sgn}(\epsilon_d^*+U^*)
\sqrt{(\epsilon_d^*+U^*)^2-D^2}\left(1-\frac{2}{\pi}
\arcsin\frac{D}{(\epsilon_d^*+U^*)}\right)\right]\quad,\\
U^*-U^0&=-\frac{2\Gamma}{D}\left[(\epsilon_d^*+U^*)
-\text{sgn}(\epsilon_d^*+U^*)\sqrt{(\epsilon_d^*+U^*)^2-D^2}\right]\quad.
\label{UrenormalizedGGD}
\end{align}
\end{widetext}

\begin{figure}[t]
  \begin{center}
    \epsfig{file=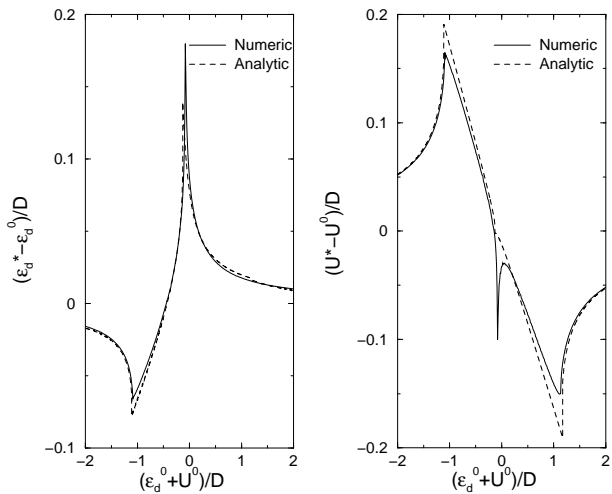,height=8cm,angle=-90}
    \caption{The relative fixed point values $\epsilon_d^*-\epsilon_d^0$ 
(left hand side) and $U^*-U^0$ 
(right hand side) as a function of the scaling variable $\epsilon_d^0+U^0$ for $V/D=0.23$, $W/D=0.17$ at $T=0$. The solid lines follow from the numerically exact solution, the dashed lines from the semi-analytical approach.}
    \label{FixedPoint}
\end{center}
\end{figure}
In Fig. \ref{FixedPoint}, the numerically exact universal functions $F$ and $G$ 
of Eqs. (\ref{UniF}) and (\ref{UniG}) are shown as functions 
of the scaling variable $\epsilon_d^0+U^0$ (solid line). 
The coupling strengths were chosen to be $V/D=0.23$ and $W/D=0.17$.\cite{FootRes} 
The semi-analytical solution (dashed line) given in Eqs. (\ref{ErenormalizedLLD} -
\ref{UrenormalizedGGD}) qualitatively agrees with the exact solution. The main difference 
is that the kink in $U^*-U^0$ near $\epsilon_d^0+U^0\approx 0$ which is present in the 
numerical solution 
is not adequately reproduced by our analytical approach, Eq. (\ref{UrenormalizedLLD}). 
This is due to the cruder decoupling scheme which does not incorportate energy-scale separation.It is interesting to note that at this point, the numerical solution yields $\epsilon_d^*+U^*=0$.
It will lead to significant differences for the phase diagram which will be discussed in the next subsection.  
 
\subsection{Critical correlation energy}
Eq. (\ref{UrenormalizedLLD}) yields
\begin{align}
U^*=\frac{U^0-2\epsilon_d^*\Gamma/D}{1+2\Gamma/D}\quad.
\end{align}
Hence, for $\epsilon_d^*>0$, 
the renormalized interaction can become negative. 

We thus define the critical initial correlation energy 
$U_c^0$ such that the renormalized correlation energy $U^*$ 
becomes zero, i.e.,
\begin{align}
U^*(U_c^0|\epsilon_d^0)\buildrel !\over=0\quad.
\end{align}
$U_c^0$ depends on the inital on-site energy $\epsilon_d^0$ and we further restrict $U_c^0$ to be positive.

The critical correlation energy $U_c^0$ is enhanced for 
positive on-site energies $\epsilon_d^0\geq0$. 
This property can most clearly be seen from the analysis of  
Eqs. (\ref{Intepsilon}) and (\ref{IntU}) 
for $T=\infty$.\cite{FootTe} Then, the Fermi function is 
constant, i.e. $f(\epsilon)=1/2$, and the integrals yield simple expressions. 
For the semi-elliptical coupling 
function, we obtain for $\epsilon_d^0<D+\Gamma/2$ 
\begin{align}
U_c^0=\frac{2\Gamma}{D-\Gamma/2}\epsilon_d^0\quad.
\end{align}

\begin{figure}[t]
  \begin{center}
    \epsfig{file=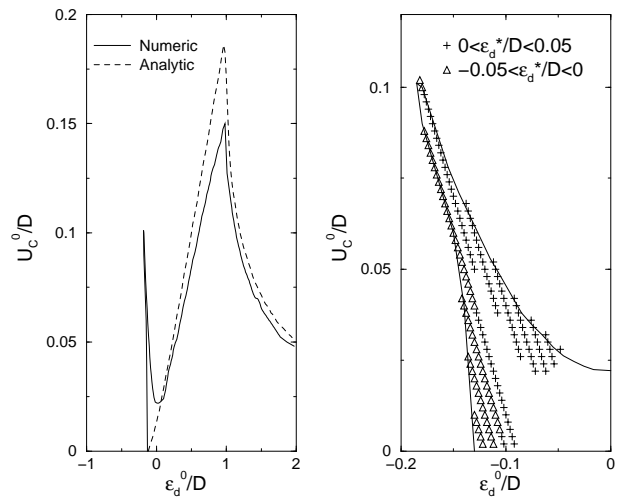,height=8cm,angle=-90}
    \caption{Left hand side: The critical correlation energy $U_c^0$ 
as function of $\epsilon_d^0$ following from the numerical (solid line) and semi-analytic (dashed line)
solution with $V/D=0.23$ and $W/D=0.17$ at $T=0$. Right hand side: 
The critical correlation energy $U_c^0$ as function of $\epsilon_d^0$ following from the numerical solution in the regime where $|\epsilon_d^*/D|\leq0.05$.}
    \label{PhaseDia}
\end{center}
\end{figure}

On the left hand side of Fig. \ref{PhaseDia}, the critical correlation energy $U_c^0$ is shown 
as a function of the initial on-site energy 
$\epsilon_d^0$ as it follows from the numerical (solid line) and semi-analytical (dashed line) solution. 
The coupling strengths were again chosen to be $V/D=0.23$ and $W/D=0.17$.  
The main difference between the two curves stems from 
the spike of $G(\epsilon_d^0+U^0)$ around $\epsilon_d^0+U^0\approx0$, seen on the right hand side of Fig. \ref{FixedPoint}. 

\subsection{Pairing correlations.}
It is now interesting to consider the possibility
of local pairing correlations.
Pair formation does not rely on particle-hole symmetry, 
in fact, asymmetries can actually enhance
this phenomenon.\cite{Tar91} But local pairing 
can only occur if - in addition to
an effective negative correlation energy 
$U^*<0$ - the renormalized on-site energy $\epsilon_d^*$
lies sufficiently close to the chemical potential 
$\mu=0$.\cite{Mar92} The local pairing 
regime is thus determined by the tradeoff between a 
high critical correlation energy $U_c^0$ and an 
effective on-site energy $\epsilon_d^*$ sufficiently 
close to the chemical potential. 

To discuss the phase diagram of this local pairing regime we will only consider the numerical solution. This approach is more reliable since it incorporates a systematic decoupling of the assisted hopping term, starting with large and ending with small energy differences with respect to $\epsilon_d^*+U^*$. The results are presented on the right hand side of Fig. \ref{PhaseDia}, where the critical correlation energy $U_c^0$ is shown as function of $\epsilon_d^0$ in the regime where $|\epsilon_d^*/D|\leq0.05$. Furthermore, we distinguish between positive ($+$) and negative ($\triangle$) renormalized on-site energy $\epsilon_d^*$.

Our analysis predicts local pairing for 
values of $\epsilon_d^0$ which favor double occupancy,
and $U_c^0 \lapp0.1$. The mean field analysis in Ref. \onlinecite{G02}
gives local pairing for $U_c^0 \lapp 0.05$ and a similar range of values of $\epsilon_d^0$
(note that we are using here comparable hybridization with the band).

It is interesting to note that the strong renormalization of $U^*$ for $\epsilon_d^0 + U^0 \approx 0$ only occurs at large values of $\ell\gg1/D^2$. Therefore, it is not present in the semi-analytical solution which does not incorporate energy-scale separation. Furthermore, there are fixed points for which $\epsilon_d^*\approx U^*\approx0$, i.e., a degeneracy between the zero-charge and two-charge state of the impurity. Since the marginal interactions also include pseudo spin-pseudo spin couplings, this might give rise to the Kondo effect in the charge channel, i.e., in the mixed valence regime.

\subsection{Marginal interactions.}
We now analyze the marginal terms, which have
been neglected so far. One term will be interpreted 
as a renormalization of the hybridization term $V_k^0$. 
This identification is crucial since the assisted hopping term 
$W_k^0$ contains an effective hybridization term which 
must be present in the fixed point Hamiltonian. 
Another term will explicitly favor local pairing. 

As in the Anderson model,
the flow generates spin-spin and pseudo spin-pseudo spin
couplings: 
\begin{align}
\label{HSW}
{\cal H}_{SW}&=-\sum_{k,k'}V_{k,k'}^*\left[\left(:\psi_k\dag\frac{1}{2}
\vec{\sigma}\psi_{k'}:+h.c.\right)\cdot
\left(\psi_d\dag\frac{1}{2}\vec{\sigma}\psi_d\right)\right.\notag\\
        &\left.-\left(:\tilde\psi_k\dag\frac{1}{2}
\vec{\sigma}\tilde\psi_{k'}:+h.c.\right)\cdot
\left(\tilde\psi_d\dag\frac{1}{2}
\vec{\sigma}\tilde\psi_d\right)\right]\quad,
\end{align}

In addition to this interaction, the following term is also generated:
\begin{align}
\label{NeglectTerms}
{\cal H}_{irr} = \frac{1}{2}
\sum_{k,k'}V_{k,k'}^*\left(:\psi_k\dag\psi_{k'}:+h.c.\right)+
\tilde\epsilon_d^*(n_d-1)
\end{align}

The existence of
these terms is independent of the assisted hopping
interaction and has been discussed in detail
in relation with the Anderson model\cite{KM94}. There, ${\cal H}_{SW}$ 
gives rise to the well-known Kondo behavior in the spin and charge sector, respectively.

${\cal H}_{irr}$ is usually neglected in the fixed point Hamiltonian 
since the first term of Eq. (\ref{NeglectTerms})
represents potential scattering 
of the conduction electrons and will be irrelevant
for our discussion. The last term renormalizes 
the on-site energy by $\tilde\epsilon_d^*\equiv\sum_kV_{k,k}^*/2$. 
For the semi-elliptical coupling function (\ref{SemiCircular}), 
we obtain $\tilde\epsilon_d^*/\epsilon_d^*\approx\Gamma/2D$ 
according to Eq. (\ref{VKK}) within the semi-analytical treatment. 
Our approximation - not to include 
newly generated coupling terms into the flow of the Hamiltonian - 
is thus consistent for $\Gamma/D\ll1$. As in Ref. \onlinecite{KM94}, we will ignore 
all terms of Eq. (\ref{NeglectTerms}) in the fixed point Hamiltonian and assume that 
they can be taking into account by redefining the
initial parameters.  

Finally, the commutator 
$[\eta,{\cal H}_{assisted}]$ also gives rise to the 
following pairing term:
\begin{align}
\label{Pair}
{\cal H}_{pair}&=
\sum_{k,k',s}W_{k,k'}^*\left[c_{k,s}\dag c_{k',-s}\dag 
d_{-s} d_s+h.c.\right]
\end{align}
which has the same shape as some of the contributions
in ${\cal H}_{SW}$, Eq. (\ref{HSW}).

The additional contributions of Eq. (\ref{HSW} - \ref{Pair}) 
need to be included in the flow equation scheme.  As these terms contain an even
number of operators associated to the localized level, but the 
generator, Eq. (\ref{generator}), contains an odd number, these
terms cannot directly change the flow equations of the parameters
$U$ and $\epsilon_d$, Eqs.(\ref{FlowD},\ref{FlowU}). They can, however, 
change the flow of the assisted hopping term, Eq. (\ref{FlowW}),
which will eventually modify Eqs.(\ref{FlowD}) and (\ref{FlowU}).
We will neglect this second order modification of the flow equations
for $U$ and $\epsilon_d$. This approximation was initially introduced in\cite{KM94} and is based on the fact that $V_{k,k'}=0$ and $W_{k,k'}=0$ for $\ell=0$.
Note that, provided that the flow equations are unchanged,
the presence of these ``marginal'' terms in the fixed point Hamiltonian
does not alter the main conclusions of the paper.

We now  approximate the magnitude of these terms at the fixed point,
neglecting their influence on the flow equations.
We obtain:
\begin{align}
\label{VKK}
&V_{k,k'}^*=2\int_0^\infty d\ell 
\eta_k(V_{k'}^0+W_{k'})\approx-
\frac{2|W_k^0V_{k'}^0|-W_k^0W_{k'}^0}{\epsilon_{k'}-\epsilon_d^*-U^*}\\
&W_{k,k'}^*=\int_0^\infty 
d\ell \eta_kW_{k'}\approx
\frac{1}{2}\frac{W_k^0W_{k'}^0}{\epsilon_{k'}-\epsilon_d^*-U^*}
\end{align}

As mentioned earlier, ${\cal H}_{SW}$ arises in the
analysis of the Anderson Hamiltonian, due to the effects
of the hybridization between the state at the impurity
and the conduction band, and can be derived
by means of a Schrieffer-Wolff transformation\cite{Sch66} . Hence,
we can interpret these fixed point interactions as arising from
an effective hopping term in the initial Hamiltonian, which, indeed,
can be obtained from a mean field decoupling of the assisted
hopping term. ${\cal H}_{pair}$ will enhance the tendency towards pair formation.

\section{Conclusions.}
We have studied a model of an impurity, or a quantum dot, with
a single electronic level, an on-site repulsion term, hybridization, and
assisted hopping. The flow equation method allows us to map
the model onto an effective Anderson model without assisted hopping,
and with additional Kondo  couplings and pair hopping terms.

The most interesting outcome of our calculation is that the
value of the on-site electron-electron interaction can become
attractive, because of the effect on the flow
of the initial assisted hopping term.
Although the Anderson model with attractive interactions
has not been as extensively studied\cite{Tar91,Mar92} as the repulsive
Anderson model, we think that our mapping suffices to give
a qualitative description of the physical properties of
the model for the interesting case of a relatively strong
assisted hopping term. Note that, in addition to
a renormalized on-site interaction term,  
terms  which describe pair hopping between the impurity
and the conduction band are generated. We expect that significant 
local pairing correlations will develop in this regime.

This result is in agreement with general arguments which suggest
that assisted hopping interactions tend to favor superconducting
ground states\cite{H93}, and with mean field studies of the same 
impurity model\cite{G02}. A related tendency towards phase
rigidity can be found in the limit where the level spacing
within the impurity is negligible with respect to the
other parameters\cite{UG91,Betal00}.

The model does not have electron-hole symmetry, and local pairing
is mainly favored when the number of electrons at the impurity
exceeds one (half filling), and when the direct hopping and
the assisted hoping terms have opposite signs, also  in agreement
with general properties of bulk systems. 
\section{Acknowledgements.}
T.S. is supported by the DAAD-Postdoctoral program. We thank J. Hirsch
for a critical reading of the manuscript.
Financial support from MCyT (Spain), through
grant no. MAT2002-04095-C02-01 is gratefully acknowledged.


\begin{thebibliography}{99}
\bibitem{FootCP}
We set the chemical potential 
equal to zero - even for finite temperatures.
\bibitem{FootTe}
The limit $T\to\infty$ is mainly academical 
since the contribution of assisted hopping 
decreases with finite temperature, i.e. 
$W\to0$ for $1/T\to0$, due to the broadening of 
the discrete spectrum of the quantum dot, see Ref. \onlinecite{G02}.
\end{thebibliography}
\end{document}